\documentclass[twocolumn,amsmath,amssymb]{revtex4}
\usepackage{graphicx}
\usepackage{dcolumn}
\usepackage{bm}

\def\tfrac#1#2{{\textstyle{#1\over#2}}}

\begin{document}

\title{The Fractional Quantum Hall Effect of Tachyons in a Topological Insulator 
Junction}
\author{Vadim M. Apalkov}
\affiliation{Department of Physics and Astronomy, Georgia State University,
Atlanta, Georgia 30303, USA}
\author{Tapash Chakraborty$^\ddag$}
\affiliation{Department of Physics and Astronomy,
University of Manitoba, Winnipeg, Canada R3T 2N2}

\date{\today}
\begin{abstract}
We have studied the tachyonic excitations in the junction of two topological
insulators in the presence of an external magnetic field. The Landau levels,
evaluated from an effective two-dimensional model for tachyons, and from the
junction states of two topological insulators, show some unique properties 
not seen in conventional electrons systems or in graphene. The $\nu=\frac13$
fractional quantum Hall effect has also a strong presence in the tachyon system.
\end{abstract}
\maketitle

The surface state of recently discovered three-dimensional topological insulators 
\cite{topo_review} contains a single Dirac cone and as a result, 
the charge carriers on the surface are characterized as massless Dirac fermions. 
Some of the properties of these particles are well known from another topological 
system, the graphene \cite{abergeletal}. We have shown earlier \cite{tachyon} 
that the dispersion relation of the surface excitations in a junction between 
two such topological insulators (TIs) show several very unique properties. 
Most importantly, under certain conditions these excitations exhibit tachyon-like 
dispersion relation \cite{tachyonics,chiao96} corresponding to superluminal 
propagation of the Dirac fermions along the interface of the two TIs. In 
addition to the tachyonic dispersion, the junction states can also support 
the usual massless relativistic dispersion relation of the Dirac fermions 
\cite{takahashi}. Here we report on the properties of these tachyonic 
excitations in an external magnetic field. We discuss the unique nature of 
the Landau levels (LLs) of these tachyons and the interaction properties 
of tachyons in a strong magnetic field, which leads to the fractional 
quantum Hall effect (FQHE) of tachyons. 

{\em Effective two-dimensional (2D) model of tachyonic excitations:} 
The tachyons in the junction of two TIs can be described by the 
effective 2D matrix Hamiltonian of the Dirac fermions but with imaginary 
Fermi velocity \cite{tachyon} (instead of the imaginary proper mass commonly
attributed to the tachyons \cite{tachyonics}, e.g., to the neutrinos \cite{chodos})
\begin{equation}
{\cal H}^{}_{\rm Tach} = \left( 
\begin{array}{cc}
\Delta^{}_0  & {\rm i} v^{}_{\rm I} p^{}_{+}  \\
{\rm i} v^{}_{\rm I} p^{}_{-} & - \Delta^{}_0     
\end{array}
\right) = \Delta^{}_0 \sigma^{}_z + {\rm i} v^{}_{\rm I} 
(\vec{\sigma} \vec{p}),
\label{Htach}
\end{equation} 
where $p^{}_{\pm} = p^{}_x \pm p^{}_y$ is the 2D momentum, $\Delta^{}_0$ is 
the effective mass of the tachyons,  ${\rm i}v^{}_{\rm I}$ is the imaginary 
Fermi velocity, and $\vec{\sigma}$ are the Pauli matrices, corresponding to 
the spin degree of freedom of an electron (tachyon). The effective Hamiltonian 
thus described has the tachyon-like dispersion, $E^{}_{\rm Tach} (p) = \pm 
\sqrt{\Delta_0^2 - v^{}_{\rm I} p^2}$, where $p=\left(p_x^2+p_y^2\right)^{\frac12}$. 
Typical values of the parameters are $\Delta^{}_0 \sim 0.3$ 
eV and $v^{}_{\rm I}\sim 10^6$ m/s. 

We subject the tachyonic system to an external magnetic field, $B$, pointing 
along the $z$-direction, i.e., perpendicular to the junction between the TIs. 
The Hamiltonian describing such a system in a magnetic field can be found from 
the Hamiltonian (\ref{Htach}) by replacing the tachyonic momentum, $(p^{}_x, 
p^{}_y)$ by the generalized momentum, $(\pi^{}_x,\pi^{}_y)$ and introducing 
the Zeeman energy, $\Delta^{}_z (B)=\frac12 g^{}_s\mu^{}_B B$. Here $\mu^{}_B$ 
is the Bohr magneton and $g^{}_s\approx 8$ is the effective $g$-factor of the
tachyonic excitations, which we assume to be equal to the $g$-factor of the
TI surface states \cite{liu_2010,wang_2010}. The effective tachyonic Hamiltonian 
in a magnetic field then takes the form 
\begin{equation}
{\cal H}^{}_{\rm Tach} (B) = \left[\Delta^{}_0 + \Delta^{}_z(B)\right] 
\sigma^{}_z + {\rm i} v^{}_{\rm I} (\vec{\sigma}\vec{\pi}).
\label{HtachB}
\end{equation} 
The LL energy spectrum corresponding to the Hamiltonian [Eq.~(\ref{HtachB})]
is characterized by an integer number $n\geq 0$ (the LL index), and is defined as
\begin{eqnarray}
E^{}_{n=0}&=&\Delta^{}_0 + \Delta^{}_z (B) \nonumber \\
E^{}_{n\geq 1,s} & = & s \left[\left[\Delta^{}_0 +\Delta^{}_z(B)\right]^2 - 
2n \left(\hbar v^{}_{\rm I}/\ell^{}_0\right)^2\right]^{\frac12} 
\nonumber \\ 
& = & s \left[\Delta^{}_0 +\Delta^{}_z(B)\right]\left[1-n
B/B^*(B)\right]^{\frac12}.
\label{En}
\end{eqnarray}
Here $s=\pm 1$, $\ell^{}_0=\sqrt{e\hbar/c B}$ is
the magnetic length, and we introduced the effective magnetic field 
$B^*(B)=\frac{e}{2\hbar c}\left[\left(\Delta^{}_0 +\Delta^{}_z(B)\right)/v^{}_{\rm I} 
\right]^2.$

The wave functions corresponding to the LLs [Eq.~(\ref{En})] can be 
expressed in terms of the functions $\phi^{}_{n,m}$, which are wave functions 
of the conventional (`non-relativistic') LLs with index $n$ and the 
intra-Landau level index, $m$, for example, the $z$ component of the angular 
momentum. The LL wave functions of the tachyons are
\begin{equation}
\Psi^{\rm (Tach)}_{n=0}=\left(\begin{array}{c}
  \phi^{}_{0,m} \\
 0
\end{array}  
 \right), \quad
\Psi^{\rm (Tach)}_{n\geq 1,s}=\left(\begin{array}{c}
 \cos \alpha^{}_{n,s}\, \phi^{}_{n,m} \\
  \sin \alpha^{}_{n,s}\, \phi^{}_{n-1,m}
\end{array}  
 \right), 
\label{fmodel2}
\end{equation} 
where $\alpha^{}_{n,s}=\sin^{-1}\left[\frac12\left(1-s \sqrt{1-n B/B^*}\right)
\right]^{1/2}.$ 

\begin{figure}
\begin{center}\includegraphics[width=9cm]{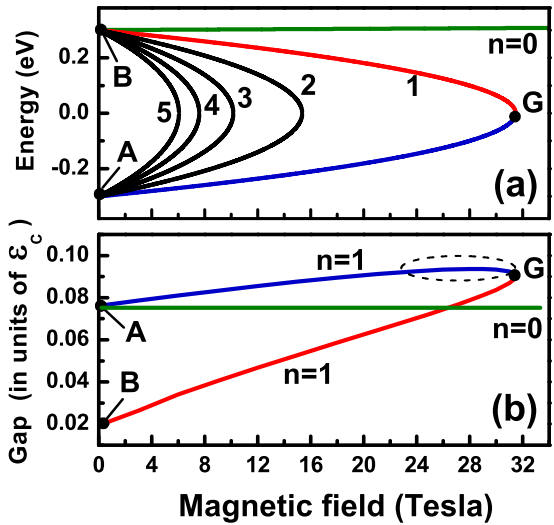}
\end{center}
\vspace*{-0.5cm}
\caption{\label{Topological1}
(a) The LLs of the effective 2D tachyonic Hamiltonian (\ref{Htach}) are
shown as a function of the magnetic field. The numbers next to the lines are 
the LL indices. The $n=1$ LL wave functions at points A and B are identical 
to conventional `nonrelativistic' LLs with indices 0 and 1, respectively.
At point G, the LL wave function of the tachyon system is identical to
the $n=1$ LL wave function of graphene. (b) $\nu =\frac13$-FQHE gap is shown 
for the $n=0$ and $n=1$ LLs of tachyons. The FQHE gap is calculated for 
a finite-size system with $N=9$ particles in a spherical geometry with parameter 
$S = 12$. The red and blue lines correspond to the $n=1$ LL branches shown 
in panel (a). The FQHE gap is measured in units of the Coulomb energy, 
$\epsilon^{}_{\rm C} = e^2/\kappa \ell^{}_0$. The dashed curve is explained
in the text.
}
\end{figure}

The LL energy spectrum obtained from Eq.~(\ref{En}) is shown in Fig.~1. The 
energy spectrum for tachyons exhibits a few distinct features: (i) The LL 
energy spectrum is mainly restricted within the energy interval $-\Delta^{}_0 
< E^{}_{n,s} < \Delta^{}_0$. (ii) At a given magnetic field $B$, only the LL 
with index $n< B^*/B$ can be observed. Therefore, for a given magnetic field, 
only a few LLs exist in the system and, for a large enough magnetic field (e.g., 
$B > 32$ Tesla in Fig.~1(a)), only the $n=0$ LL survives. This behavior of the 
LL of tachyonic excitations is totally different from that in
the `non-relativistic' semiconductor systems with the LL energy 
spectrum $E^{}_n \propto nB$, and ``relativistic" graphene system with the LL 
energy spectrum $E^{}_n \propto \sqrt{nB}$ \cite{abergeletal}. However, there 
is one similarity between the tachyonic LL dispersion relations and that
in graphene. In both cases there is one $n=0$ LL, the energy of 
which is independent of the strength of the magnetic field (without the Zeeman 
splitting). In graphene, the energy of this LL is $E^{}_{n=0}=0$, while for 
the tachyonic excitations, $E^{}_{n=0}=\Delta^{}_0$. In both cases the 
corresponding wave functions are $\phi^{}_{n=0,m}$. 

To address the similarities and differences between the LL wave functions 
of the tachyonic system and those of graphene (or even conventional 
semiconductor systems), we consider the interaction properties of tachyonic 
excitations in a given LL. To characterize the strength of the inter-tachyonic 
interactions we study the strength of the FQHE, 
i.e., the magnitude of the FQHE gap. In the FQHE regime the electrons partially 
occupy a single LL and the properties of such a system are characterized
by the inter-particle interactions within the corresponding LL \cite{FQHE_book}. 
The interaction strength within a given LL is determined from the Haldane 
pseudopotentials, $V^{}_m$ \cite{Haldane_83}, which are the interaction energies 
of two particles with relative angular momentum $m$. The pseudopotentials are 
determined from \cite{Haldane_83}
\begin{equation}
V_m^{(n)} = \int_0^{\infty} \frac{dq}{2\pi} q V(q) 
\left[F^{}_n(q) \right]^2 L^{}_m (q^2) 
 e^{-q^2}, 
\label{Vm}
\end{equation}
where $L^{}_m(x)$ are the Laguerre polinomials, $V(q) = 2\pi e^2/(\kappa \ell^{}_0 
q)$ is the Coulomb interaction in the momentum space, $\kappa$ is the dielectric 
constant, and $F^{}_n(q)$ is the form factor for the $n$-th Landau 
level. The form factor is determined by the structure of the LL wave functions. 
For the tachyonic system the form factor is given by 
\begin{eqnarray}
& & F^{}_{n=0}(q) = L^{}_0\left(\tfrac12q^2 \right)  \label{f0}  \\
& & F^{}_{n\geq 1}(q) =\cos^2\alpha^{}_{n,s} L^{}_n \left(\tfrac12q^2\right) 
+\sin^2\alpha^{}_{n,s} L^{}_{n-1} \left(\tfrac12q^2\right).
\label{fn}
\end{eqnarray}
The form factor of the $n=0$ LL [Eq.~(\ref{f0})] is identical to that of 
graphene and also to that of the non-relativistic 
systems. However, for $n\geq 1$ the form factor of the tachyonic system becomes 
unique. For graphene and for the non-relativistic systems the 
corresponding form factors are $F_n^{\rm (NR)} = L^{}_n$ and $F_n^{\rm (Gr)} = 
\frac12 (L^{}_n + L^{}_{n-1})$, respectively. In both cases the form factors are 
independent of the magnetic field. For the tachyonic system, on the other hand,
the form factor (\ref{fn}) depends on the magnetic field, through the
effective angle $\alpha^{}_{n,s} (B)$. With increasing magnetic field the tachyonic 
form factor, $F^{}_{n\geq 1}$, changes from the non-relativistic value, $F_n^{(NR)}$ 
[point B in Fig.~1(a)] or $F_{n-1}^{(NR)}$ [point A in Fig.~1(a)], in a small 
magnetic field, $B\rightarrow 0$, to the form factor of graphene, $F_n^{\rm (Gr)}=
\frac12 (L^{}_n + L^{}_{n-1})$ \cite{Apalkov_06}, for $B=B^*$.  

\begin{figure}
\begin{center}\includegraphics[width=9cm]{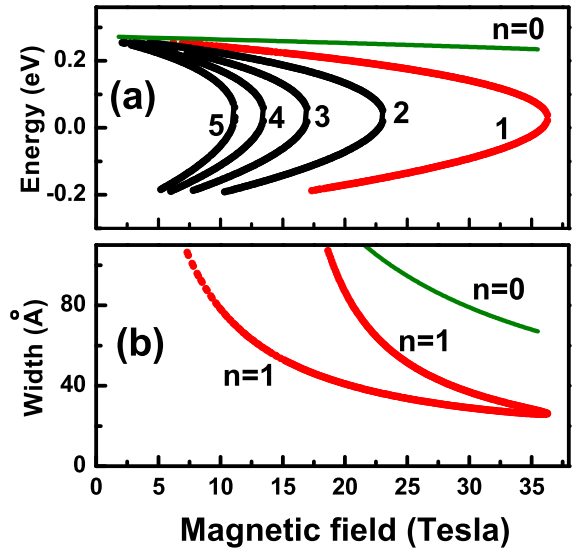}
\end{center}
\vspace*{-0.5cm}
\caption{\label{Topological2}
(a) The LL energy spectrum of a electron in a junction between two TIs
is shown as a function of magnetic field
for a few lowest LLs. The numbers next to the lines are the LL indexes.
The TIs have the same Hamiltonian parameters, except $A^{}_1$, which
takes the values: $A^{}_1 = 2.2$ eV$\cdot$\AA\ for TI-1 and $A^{}_1 = 4.0$ 
eV$\cdot$\AA\ for TI-2. (b) The width in the $z$-direction of the LL wave functions 
is shown for $n=0$ and $n=1$ LLs. The corresponding LL energy spectrum is shown 
in panel (a).
}
\end{figure}

The FQHE with an incompressible ground state can be observed only 
in the LL with strong short-range repulsion, i.e., a fast decay of 
the corresponding pseudopotentials, $V^{}_m$, with $m$. Such a strong 
repulsion is realized only in the LL with a strong admixture of 
$\phi^{}_{0,m}$. Therefore, in a tachyonic system the FQHE is expected 
only in the $n=0$ and $n=1$ LLs. To study the strength of the corresponding 
FQHE we numerically evaluate the energy spectrum of a finite $N$-electron 
system in a spherical geometry \cite{Haldane_83} with the radius of the sphere 
$\sqrt{S}\ell^{}_0$. Here $2S$ is the integer number of magnetic fluxes through 
the sphere in units of the flux quantum. For a given number of electrons, the 
radius of the sphere determines the filling factor of the system. For example, 
the $\nu=1/q$ FQHE ($q$ is an odd integer) corresponds to the relation 
$S=(\frac{q}{2})(N-1)$ \cite{FQHE_book}. 

In Fig.~1(b) the $\nu = \frac13$-FQHE gap is shown for $n=0$ and $n=1$ LLs of 
the tachyonis system. For the $n=0$ LL, the FQHE gap does not depend on $B$ 
and the gap exactly equals to the FQHE gap for the $n=0$ non-relativistic LL. 
This is because the $n=0$ tachyonic LL wave function consists of only the 
functions $\phi^{}_{0,m}$. For the $n=1$ LL, the 
wave function is the $B$-dependent mixture of $\phi^{}_{0,m}$ and 
$\phi^{}_{1,m}$. As a result the FQHE gap depends on the magnetic 
field and changes from the $n=0$ non-relativistic LL value for $B\rightarrow 0$ 
(point A) to the $n=1$ graphene LL value for $B= B^*$ (point G) and finally to 
$n=1$ non-relativistic LL value at point B (in the thermodynamic limit 
such a state becomes compressible). 

The maximum FQHE gap in a non-relativistic system corresponds to the green line 
in Fig.~1(b), while the maximum FQHE gap in the graphene system corresponds to 
point G in Fig.~1(b). Therefore, comparing the data in Fig.~1(b), we 
conclude that within some range of the magnetic fields (which is shown 
in Fig.~1(b) by an oval curve), the FQHE gap in the model tachyonic system is the 
largest compared to other available 2D systems. The tachyonic dispersion relation 
provides an unique possibility to study, within a single tachyonic $n=1$ LL, the 
properties of the $n=0$ non-relativistic LL [point A in Fig.~1(a)], $n=1$ 
graphene LL [point G in Fig.~1(a)], and $n=1$ non-relativistic LL [point B 
in Fig.~1(a)].

{\em Three-dimensional (3D) model for the junction states between two TIs:}
Until now, we have discussed the magnetic field effects via an effective Hamiltonian
for tachyons. The tachyonic dispersion can be realized in the junction of two TIs 
\cite{tachyon}. The junction dispersion relation in this case is approximately 
described by the effective Hamiltonian (\ref{Htach}). The realization of the 
tachyonic excitations as the junction states bring additional factors into 
consideration. For example, the junction states have a finite width in 
the $z$-direction which can reduce the interaction strength in that system. 

We consider the junction between two TI insulators: TI-1 for $z<0$ and TI-2 for $z>0$. 
Here $z=0$ corresponds to the junction surface. The electronic properties of both 
TIs are described by the same type of low-energy effective 3D Hamiltonian
\cite{liu_2010,zhang_2009} of the matrix form
\begin{equation}
{\cal H}^{}_{\rm TI}=\epsilon(\vec{k})+ \left(  
\begin{array}{cc}
 M(\vec{p})\sigma^{}_z- {\rm i} A^{}_1\sigma^{}_x \partial^{}_z & (A^{}_2/\hbar) 
p^{}_{-}\sigma^{}_x \\
(A^{}_2/\hbar) p^{}_{+}\sigma^{}_x & M(\vec{p})\sigma^{}_z + {\rm i} 
A^{}_1 \sigma^{}_x \partial^{}_z  
\end{array}
\right),
\label{HTI}
\end{equation}
where  $\partial^{}_z = \partial/\partial z$,  and 
\begin{eqnarray}
& & \epsilon (\vec{k})=C^{}_1 - D^{}_1\partial_z^2+(D^{}_2/\hbar^2 )(p_x^2 + p_y^2),\\ 
& & M(\vec{k})= M^{}_0 + B^{}_1 \partial_z^2 - (B^{}_2/\hbar^2 ) (p_x^2 + p_y^2).
\end{eqnarray}
We assume that for TI-1 the constants in the above Hamiltonian are the same 
as for $\mbox{Bi}^{}_2\mbox{Se}^{}_3$ \cite{zhang_2009}, while the for TI-2 the 
constants are different but close to the values for $\mbox{Bi}^{}_2\mbox{Se}^{}_3$. 
We assume that only the constant $A^{}_1$ is different for the two TIs, i.e., 
$A^{}_1 = 2.2$ eV$\cdot$\AA\ for TI-1 and $A^{}_1 = 4.0$ eV$\cdot$\AA\ for TI-2. 
All other constants in the Hamiltonian (\ref{HTI}) are kept the same for both TIs
\cite{tachyon}. For these parameters, the junction states exhibit the tachyonic 
dispersion \cite{tachyon}. The four-component wave function corresponding to the 
Hamiltonian (\ref{HTI}) determines the amplitudes of the wave functions at the 
positions of Bi and Se atoms: $(\mbox{Bi}^{}_{\uparrow}, \mbox{Se}^{}_{\uparrow}, 
\mbox{Bi}^{}_{\downarrow}, \mbox{Se}^{}_{\downarrow})$, where the arrows indicate 
the direction of the electron spin. 

The Hamiltonian of the TI in an external magnetic field, pointing along the
$z$-direction, can be obtained from the  Hamiltonian (\ref{HTI}) by replacing the
2D momentum $(p^{}_x,p^{}_y)$ by the generalized momentum $(\pi^{}_x, \pi^{}_y )$ 
\cite{yang_2011} and introducing the Zeeman energy, $\Delta^{}_z=\frac12 g^{}_s
\mu^{}_B B$. For the Hamiltonian (\ref{HTI}) in a magnetic field, the wave function 
in the $n$-th LL has the general form 
\begin{equation}
\Psi^{\rm (TI)}_{n\geq 1} (z)  = 
\left( \begin{array}{c}
 \chi^{(n)}_1 (z) \phi^{}_{n-1,m} \\
  \chi^{(n)}_2 (z)   \phi ^{}_{n-1,m} \\  
  \chi^{(n)}_3 (z)  \phi^{}_{n,m} \\
  \chi^{(n)}_4(z) \phi ^{}_{n,m}  
\end{array}  
 \right).
\label{fTI}
\end{equation}
Therefore, just as for the tachyonic states the wave function $\Psi^{\rm 
(TI)}_{n\geq 1}$ is the mixture of $n$ and $n-1$ non-relativistic LL functions. 
For the $n=0$ LL, only $\chi^{(n=0)}_3$ and $\chi^{(n=0)}_{4}$ are non-zero. 
 
To find the LL junction states we follow the same procedure as for the LL surface 
states of a TI \cite{zhou_2008,shan_2010,yang_2011}. For each TI we find 
the general bulk solution of the Schr\"{o}dinger equation in the form of $\Psi 
\propto e^{\lambda^{(m)} z}$, where $\lambda^{(m)}$ is a complex constant, and  
$m=1$ and 2 for TI-1 and TI-2, respectively. This solution has a given energy, 
$E$, and a given LL index, $n$. The corresponding $\lambda^{(m)}$ are determined 
from a secular equation, $\det \left[ {\cal{H}}^{(m)}_{TI} (\vec{k}, \lambda^{(m)}) 
- E\right] = 0$, for each TI. For each energy $E$ and the LL index $n$, the 
secular equation provides eight values of $\lambda^{(m)}_j(n, E)$, 
$j = 1, \ldots, 8$ with the corresponding wave functions. Second, since we 
are looking for the localized LL junction states, we need to choose (for each TI) 
only four values of $\lambda ^{(m)}_j$ out of eight with the
properties: $\mbox{Re} \lambda ^{(m)}_j > 0 $  for TI-1 ($z<0$)  and 
$\mbox{Re} \lambda^{(m)}_j <0$ for TI-2 ($z>0$). We then choose the 
corresponding four wave functions (for each TI) as the basis and expand the solution 
for the LL junction state in this basis. Finally, the energy of the LL junction 
state is found from the condition of continuity of the wave function, $\Psi ^{\rm 
(TI)}_{n}(z)$, and current $[\delta {\cal H}^{}_{\rm TI} / \delta k^{}_z] 
\Psi ^{\rm (TI)}_n$ in the junction. 

\begin{figure}
\begin{center}\includegraphics[width=9cm]{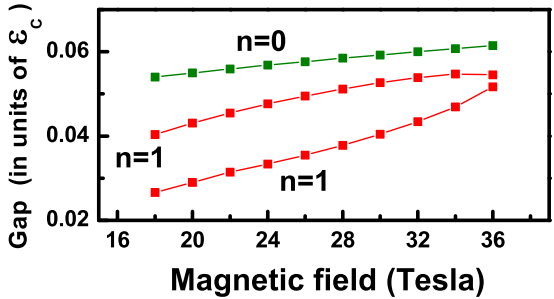}
\end{center}
\vspace*{-2.5cm}
\caption{\label{Topological3}
The $\nu =\frac13$ FQHE gap for $n=0$ (green line) and $n=1$ (red line) LLs in the
junction of two TIs.  The corresponding LLs are shown in Fig.~2(a) by green and
red lines, respectively. The FQHE gap is calculated for a finite-size system with 
$N=9$ particles in spherical geometry with parameter $S = 12$. The FQHE gap is
measured in units of Coulomb energy, $\epsilon^{}_{\rm C} = e^2 /\kappa \ell^{}_0$.
}
\end{figure}

The spectrum of the LL junction states corresponding to the tachyonic dispersion 
is shown in Fig.~2(a). The spectrum is qualitatively similar to that
[see Fig.~1(a)] obtained from the model 2D Hamiltonian. Both spectra have the 
finite range of magnetic fields and energies, where the LLs can be observed. 
At weak magnetic fields, the difference between the LL spectrum of the junction states 
and the 2D model is clearly visible. For a given LL index $n$, there are 
no junction states for weak magnetic fields. These junction states are delocalized 
in the $z$-direction. To illustrate this delocalization we show in Fig.~2(b) the width 
of the $n=0$ and $n=1$ LL wave functions in the $z$-direction. At a singular point of 
the LL spectrum, i.e., for $B=B^*$, where the derivative of the LL energy with the magnetic 
field becomes infinitely large, the LL wave functions have the smallest width. This 
width increases with decreasing magnetic field and finally the LL junction states 
are delocalized in the $z$-direction. A similar behavior is observed for $n=0$ LL, 
but there are no singular point in this case. Therefore, the LL energy spectrum 
of the junction states in the regime of tachyonic dispersion can be well described 
within 2D effective model near the singular point $B\sim B^*$.


We have evaluated the FQHE gaps for the $n=0$ and $n=1$ junction LLs. We have 
used the same approach as for the 2D model of the tachyonic excitations, discussed 
above. The results are shown in Fig.~3. Quantitatively the 
behavior of the FQHE gap as a function of the magnetic field is similar to 
the 2D model of the tachyonic excitations [Fig.~1(b)]. Due to a finite width 
of the LL wave functions in the TI junction, there is a reduction
of the inter-electron interaction strength and correspondingly the FQHE gap. This 
reduction is visible for $n=0$ LL, where a smaller FQHE gap and the magnetic field
dependence of the FQHE gap is shown in Fig.~3. 

Although for the 2D tachyonic model the FQHE gap is the largest for the 
$n=1$ LL, for the junction LLs the FQHE gap is the largest for the $n=0$ 
LL, due to the non-zero spin polarization of the $n=1$ junction LL.
This spin polarization is clearly visible from the general property of 
the LL wave function [Eq.~(\ref{fTI})]; only the components $\chi_3^{(n=0)}$ 
and $\chi_4^{(n=0)}$ of $\Psi_{n=0}^{\rm (TI)}$ are non-zero and these 
components correspond to the spin-down polarization. The numerically found $n=1$ LL 
wave functions also show partial spin-down polarization. As a result, the LL wave 
function have larger contribution from the $\phi^{}_{n=1}$ non-relativistic LL 
function, which reduces the inter-electron interaction strength and the FQHE gap. 

To summarize: we have investigated the magnetic field effects of tachyonic
excitations along the interface of two topological insulators. We used an
effective two-dimensional model Hamiltonian for tachyons and the
three-dimensional model for the junction states of the two TIs, both developed
by us \cite{tachyon}. The Landau levels
in both these models show very similar behaviors. Unlike in graphene or in
conventional electron systems, only a few LLs are found to exist for the tachyons.
Only one LL ($n=0$) survives for large magnetic fields. The $\nu=\frac13$
FQHE is the strongest (within a limited range of the magnetic field) when compared
with that for conventional electron systems and graphene. Interestingly, the
FQHE in the $n=1$ LLs for tachyons describes the FQHE of the $n=0, 1$ LLs of the
non-relativistic electron system and that of the $n=1$ graphene LL
in different regions of the magnetic field. Experimental confirmation of these
properties of the Landau levels would provide strong evidence on the existence 
of elusive tachyons. 

The work has been supported by the Canada Research Chairs Program of the
Government of Canada.

\end{document}